  \documentstyle[aps,prl] {revtex}

\begin {document}

  \title {First-Principles Studies of Hydrogenated Si(111)--7$\times$7}

  \author { D.R. Alfonso$^{a}$, C. Noguez$^{b,c}$, D.A. Drabold$^b$
   and S.E. Ulloa$^b$ }

  \address { $^a$Department of Chemistry, $^b$Department of Physics and
   Astronomy, and $^{a,b}$Condensed Matter \\ 
   and Surface Sciences Program, Ohio University, Athens, Ohio
   45701--2979  \\
   $^c$Instituto de F\'{\i}sica, Universidad Nacional Aut\'onoma de 
   M\'exico, Apdo. Postal 20-364, 01000 M\'exico D.F., M\'exico }

  \date {19 October 1995}

  \twocolumn

  \maketitle

  \begin {abstract} 
  
The relaxed geometries and electronic properties of the hydrogenated
phases of the Si(111)-7$\times$7 surface are studied using
first-principles molecular dynamics. A monohydride phase, with one H
per dangling bond adsorbed on the bare surface is found to be
energetically favorable.  Another phase where 43 hydrogens saturate the
dangling bonds created by the removal of the adatoms from the clean
surface is found to be nearly equivalent energetically. Experimental
STM and differential reflectance characteristics of the hydrogenated
surfaces agree well with the calculated features.

  \end   {abstract}


  \narrowtext

The identification of the fascinating 7$\times$7 reconstruction on the
(111) surface of silicon has motivated continued interest in its
structural and electronic properties over the years.  The
Si(111)-7$\times$7 surface is perhaps the system that best
exemplifies the important interplay between structure and electronic
properties in surface physics.  The work of Binnig {\em et
al.} using scanning tunneling microscopy (STM) \cite{binning} led
Takayanagi {\it et al.} to propose their
dimer-adatom-stacking fault (DAS) model whose structure is consistent
with many experiments \cite{taka}. The geometric and electronic
properties of the 7$\times$7 surface are well known through a
combination of experiments \cite{binning,taka,demuth,wolkow,layet} and 
realistic calculations ---
especially the first-principles investigations of Stich {\em
et al.} \cite{stich} and Brommer {\it et al.} \cite{brommer} using
large supercells in state-of-the-art approaches.

The structure of the hydrogenated phases of this important surface are
not as well understood. Studies of this surface reaction and
associated reconstruction are of technological relevance in the
chemisorption processes of H, NH$_2$, \cite{wolkow}
and hydrocarbons \cite{pian}.  In addition, the hydrogenated surface
can be used as reference in clarifying the optical response of the bare
7$\times$7 substrate, and is then important to have a better
understanding of its structural and dynamical properties
\cite{noguez,roy}.  From a computational standpoint, 
Si(111)-7$\times$7:H surfaces are challenging 
due to the large size unit cells.

Experimental evidence suggests that adsorption of hydrogen on
Si(111)-7$\times$7 occurs in a two-step process
\cite{roy,mortensen,karlsson}.  The first step likely occurs when only
the 19 dangling bonds (DBs) per unit cell are saturated, corresponding
to those of the 12 adatoms (ADs), 6 rest atoms (RAs) and corner hole
atom (CH) of the DAS model, to form Si(111)-7$\times$7:19H.\@  The
second step occurs when the adatom layer is missing and hydrogen
saturates the corresponding 43 DBs to form Si(111)-7$\times$7:43H
\cite{karlsson}.  These findings have resulted in several unresolved
issues, such as whether the DAS structure remains unchanged when the
surface is terminated by H atoms, which of the two hydrogenated phases
is more stable, and what are the main electronic properties of the
Si(111)-7$\times$7 surface after passivation with hydrogen.  To date,
the efforts to address these issues theoretically have been limited to
different reconstructions of this surface \cite{noguez,roy}, and
first-principles studies of the surface upon H and NH$_2$ adsorption
(using a cluster model for the substrate) \cite{ye}.  An alternative
and more reliable approach is to perform molecular dynamics supercell
calculations that realistically model the surface using the full unit
cell of the hydrogenated phase, combined with the accuracy of an {\em
ab initio} treatment of the interatomic interactions. 

Here, we present the first {\em ab initio} studies to
characterize the Si(111)-7$\times$7:19H and :43H systems using large
supercell geometries approaching 400 atoms.  Application of this
method to elucidate the structural and electronic properties of the
clean 7$\times$7 surfaces yields results in good agreement with
previous work.  We find that H adsorption on the clean surfaces
leads preferentially to a monohydride structure
(Si(111)-7$\times$7:19H), and is accompanied by considerable surface
relaxation, in agreement with experiments.  The other hydrogenated
phase identified, the Si(111)-7$\times$7:43H surface, is found to
approach energetically the :19H phase, explaining the possible
coexistence reported in experiments.  Moreover, we find that the
electronic density of states of these H-passivated surfaces exhibit no
surface states in the gap, while still showing strong contributions
from back-bonding states at low H-coverage, in agreement with STM and
differential reflectance measurements \cite{roy,mortensen}.

Our computational scheme is molecular dynamics simulations based on
density functional theory in the local density approximation, using the
Harris functional, Hamann-Schl\"uter-Chiang pseudopotentials
\cite{HSC}, and a minimal $s$-$p$ basis representation.  [A
comprehensive discussion of this scheme, approximations used, and a
description of various tests can be found in Ref.\ \ref{sankey}.]
In silicon, the $\pi$-bonded (2$\times$1) reconstruction of the (111)
surface, the 2$\times$1, p(2$\times$2) and c(4$\times$2) phases of the
(100) surfaces, as well as the 5$\times$5 DAS structure of Si(111) have
all been studied by this technique \cite{adams}.  

As a test, we started applying this approach to investigate the
structural properties of clean Si(111)-7$\times$7.  The substrate is
represented by a supercell with ten Si-layers plus the adatoms on the
slab surface (the initial coordinates of adatoms and the three
uppermost Si layers were those of Ref.\ \ref{brommer}).  The model
system contains all the features of the DAS configuration (see Fig.\
\ref{si7x7}a).  Periodic boundary conditions are imposed parallel to
the surface, and the bottom Si layer is terminated by H atoms to
passivate dangling bonds.  The equilibrium structure of the 543 atom
slab is then obtained  with a dynamical quenching minimization
technique.  All the atoms, except for the bottom Si layer, were allowed
to relax until the forces on them converged to $< 0.1$ eV/\AA.  Only
the $\Gamma$ point of the small Brillouin zone of the supercell is used
to sample the electronic states. The calculated bondlengths of ADs
with the first layer atoms are found to 
be stretched from 3 to 6\% with respect to a typical Si-Si bond length
(2.35 \AA). The average distance between the AD and the second layer
atoms directly beneath them is about 2.44 \AA, except for two adatoms
on the faulted region of the unit cell (shown as 1 and 2 in Fig.\
\ref{si7x7}a), which have a distance 8\% larger than the typical 
value. The bondlengths between RAs and second layer atoms are stretched
by 4\%, as they are 0.5 \AA\@ away from the first layer.  The relative
heights of AD 3 to 6 are 0.069, 0.055, 0.0 and 0.025 \AA, respectively.
These values are in qualitative agreement with those from LEED, 0.12,
0.08, 0.0 and 0.04 \AA \cite{leed}.  The second and third layer atoms
directly below ADs move $\sim$0.5 \AA\@ towards the bulk, and their
bondlengths are $\sim2$\% shorter than the bulk Si-Si bondlength. 
Comparison of our relaxed atomic positions with those from
Car-Parrinello work \cite{brommer} show nearly identical geometry
\cite{comment}.  We adopted these equilibrated positions for the
geometry of the clean surface, as it is in general agreement with
values reported in previous work \cite{brommer,leed,comment,north}, and
then proceed to saturate dangling bonds.

The Si(111)-7$\times$7:19H is modeled by terminating all the silicon
DBs of the bare surface with H atoms (see Fig.\ \ref{si7x7}b), and all
atoms --- except for the bottom Si layer --- of the 366-atom supercell
are allowed to relax (for computational speed, we take only six Si
layers here, plus adatoms and chemisorbed hydrogens) \cite{comment2}.  
The positions of the surface Si atoms change
when the 19 DBs are terminated by H atoms, and the heights of ADs
become nearly similar.  Relative to the clean surface, adatoms and
first layer atoms move away from the bulk layer, while RAs, CH and ADs
1 and 2 move towards the slab, leaving a nearly flat top layer
\cite{comment}.  The ADs BB lengths with first layer atoms are slightly
shorter ($\sim 1$\%) than in the clean substrate, while those of the RA
with second layer atoms are comparable to the bare surface. The
calculated Si-H bond distances are 1.54 \AA, nearly 8\% larger than the
sum of the covalent radii of Si and H (1.43 \AA), and comparable to the
value obtained from other calculations (1.502--1.520) \cite{ye}.  The
dimer atoms and second and third layer atoms directly below ADs remain
in their clean-surface positions.  Our calculation predicts an energy
of $-0.27$ eV/(unit cell area) for the hydrogen-covered surface,
relative to clean substrate plus gas-phase H$_2$ molecules.  This
indicates that the formation of the monohydride phase on
Si(111)-7$\times$7 is stable, in agreement with experiments
\cite{roy,mortensen,karlsson,sakurai}.  

The Si(111)-7$\times$7:43H surface has been suggested to co-exist with
:19H, based on STM observations \cite{mortensen}. The first two layers
of the surface atoms in :43H (including surface H), both on the
faulted and unfaulted half of the unit cell, are found to resemble the
hydrogenated bulk terminated Si(111)-1$\times$1 surface.  As in the
:19H monohydride phase, dimers on the second layer and a corner hole
atom are also present (see Fig.\ \ref{si7x7}c).  Although AD removal
is found to have a considerable activation barrier, it is easily
compensated by the formation of Si-H bonds \cite{mortensen}.  This
surface is modeled by removing the ADs from the clean slab model and
passivating the resulting 43 DBs with H, allowing this 378-atom
supercell to relax.  The equilibrated structure shows Si-H bonds
directed normal to the surface and bondlengths comparable to the :19H
phase.  The uppermost Si atoms are three-fold coordinated to the atoms
directly below, with bondlengths stretched by 1 to 2\% relative to the
normal Si-Si bond, comparing well with LDA results of the hydrogenated
1$\times$1 surface \cite{ye}.  The RAs and CH move towards the bulk by
about 0.5 \AA, respect to the clean surface, while the second and
third layer atoms, originally directly below ADs, move away from the
slab and reach their typical bulk positions. The
Si(111)-7$\times$7:43H phase is found to be merely 0.12 eV/(Si-H bond)
less stable than the :19H monohydride phase, indicating that these two
structures are almost energetically equivalent.  This result suggests
the energetic may indeed allow these two phases to coexist, as
reported in Ref.\ \ref{mortensen}.
 
The calculated electronic density of states (EDOS) also contains
important information on these systems.  The clean Si(111)-7$\times$7
surface contains midgap states which we can clearly identify with
surface dangling bonds (see Fig.\ \ref{edos}).  These EDOS features,
lying above 0.19 eV and from 0 to $-0.29$ eV, arise from the empty and
filled AD DBs, respectively. [Energies are referred to the Fermi level,
which is 0.64 eV above the valence band edge $E_v$, in close agreement
with the experimental value of 0.6 eV in Ref.\ \ref{demuth}.]   The
dangling bond states are localized at various sites with a predominant
$p_z$ character. Between $-0.88$ and $-0.98$ eV, we find states
resonant with the valence band which involve rest atom DBs, while AD
backbonds lie between $-1.38$ and $-1.44$ eV.  Corner hole states
appear at $-1.08$ and $-1.24$ eV.\@  Our results are in good agreement
with several experiments that have reported occupied AD and RA dangling
bonds at $-0.25$, and $-0.85$;  AD backbonds and corner hole states at
$-1.8$ eV; and AD back-antibonds at 0.55 eV \cite{layet}.

Figures \ref{stm}a--c exhibit simulated STM images according to the
scheme of Tersoff and Hamann \cite{tersoff}, with the electronic states
obtained in our calculation.  The panels display constant charge
density plots ($\rho \sim$ 0.01 eV/\AA$^3$) showing surface states at
various energies below the Fermi level.   Over a wide range of low bias
values, for energies close to the Fermi level, we see a clear asymmetry
between the faulted and unfaulted portions of the unit cell (not shown
here), in agreement with low-bias STM measurements \cite{layet}.  
Figure \ref{stm}a exhibits states between $-0.88$ and $-1.08$ eV,
corresponding to DBs of RA and CH, while Fig.\ \ref{stm}b exhibits BB
states lying between $-1.38$ and $-1.40$ eV, arising from $p_x$ and
$p_y$ orbitals of the AD bonded to $p_z$ orbitals of the first layer
atoms.  The general features of our calculated STM images for various
RA/CH and AD are in excellent agreement with those obtained by Hammers
{\it et al.} \cite{layet}.  

In marked contrast to the clean surface, the EDOS of
Si(111)-7$\times$7:19H has no midgap states, as shown in Fig.\
\ref{edos}.  This is intuitively expected, as the DBs responsible for
these states in the bare surface are passivated by H atoms.   This
finding corroborates EDOS measurements in Ref.\ \ref{mortensen} and
\ref{sakurai} where features due to DBs on the bare surface disappear
as the substrate is dosed with hydrogen.  [Also, theoretical STM images
of states near $E_v$ yield featureless isodensity plots.]   Moreover,
we find that states lying between $-0.74$ eV and $E_v$ are extended and
show no contribution from the surface H and ADs states.  Notice,
however, that backbond states associated with the surface adatoms are
still present, and lie between 1.07 and 1.21 eV below the Fermi level,
and yield the STM image shown in Fig.\ \ref{stm}c.  This finding
supports interpretation of recent differential reflectance experiments
\cite{roy}, as discussed below \cite{noguez}.  

The EDOS of the Si(111)-7$\times$7:43H also contains no midgap states
(see Fig.\ \ref{edos}).  A clear reduction of the density of states in
the region below $E_v$ and in the conduction band is produced by the
disappearance of ADs states. [Notice in particular the absence of a
sharp feature at $\sim -1.6$ eV.]  Our results agree quite well with
the local tunneling spectroscopy experiment of Mortensen {\em et al.}
\cite{mortensen}, who find no high density of states near the Fermi
level for the :43H-coverage.  Only extended states due to surface and
bulk Si were found between $-0.8$ eV and the valence band edge $E_v$,
as all the dangling bonds are fully saturated.

Our findings also provide strong support for the interpretation of
differential reflectance (DR) measurements by Roy and Borensztein
\cite{roy}, who employ this powerful technique to explore the evolution
of the surface electronic states under various chemisorption processes.
In particular, at zero and low H-coverage, a strong DR feature at 1.6
eV is identified as coming from transitions involving the backbond
states of adatoms, as supported by tight-binding calculations of model
structures \cite{noguez}.  As the H-coverage increases, however, this
feature saturates and higher energy transitions dominate the DR
spectra.  This behavior is consistent with the disappearance of the
backbond states as the higher coverage is reached and the :43H phase
forms.  Notice that the saturation of the 1.6 eV feature is accompanied
in experiments by a strong reconstruction of the surface, as probed by
LEED, and as expected from the relaxation pattern shown in Fig.\
\ref{si7x7}c \cite{karlsson}.  Finally, we should also notice that an
empirical-parameter calculation of the optical response of this surface
using our fully relaxed atomic positions yields excellent agreement
with the DR experiments, as will be discussed elsewhere
\cite{cecilia-ya-mero}.

In summary, we obtained significant geometrical relaxation arising
from H-adsorption on the clean 7$\times$7 surface, accompanied
by strong shifts of the character and energies of the electronic
states.  These changes are found to be in excellent agreement with STM,
differential reflectance, and LEED experiments, and provide the basis
for detailed comparisons with new experimental measurements.

We are thankful to Y. Borensztein, R. Del Sole, P. Ordej\'on,
and R. G. Barrera for valuable comments and cooperation.
C.N.\@ is partly supported by the University of Mexico grants
DGAPA-IN-102493 and PADEP-003309, D.A.D.\@ is partly supported by NSF
grant No. DMR-93-22412, and S.E.U.\@ acknowledges the support of the
A.\@ von Humboldt Foundation.

 \begin{figure}
 \caption{Top view of (a) the DAS model for the Si(111)-7$\times$7 bare
surface (big gray, black and small gray circles represent adatoms (AD), 
rest atoms (RA) and corner hole atom (CH), and first layer atoms, 
respectively); (b) the :19H coverage (small white circles denotes H
atoms); and (c) the :43H coverage surface.
The faulted portion of the outlined unit cell is on the
left. Unit cell shown by dashed lines.}  
 \label{si7x7} 
 \end{figure}

 \begin{figure} 
 \caption{ Electronic density of states for the clean
Si(111)-7$\times$7 surface, as well as for the two hydrogenated
surfaces studied with 19 and 43 H-atoms per cell.\@  All energies
referred to the Fermi level; level broadening used 0.1 eV; curves
displaced vertically for clarity.}  
 \label{edos}
 \end{figure}

 \begin{figure} 
 \caption{Simulated STM images in the Si(111)-7$\times$7 system. (a)
RA/CH levels (at $\sim -1$ eV in Fig.\ 2), and (b) AD backbond states
($\sim -1.4$ eV) for the clean surface.  (c) AD backbond states in the
monohydride :19H phase ($\sim -1.2$ eV in Fig.\ 2). The faulted portion
of the outlined unit cell is on the right.}
 \label{stm}
 \end{figure}
 

\begin{references}
\bibitem{comment} The only large difference between our bare surface
results and those of Brommer's involves the heights of adatoms 1 and 2
on the faulted region of the supercell.  These adatoms are displaced
towards the vacuum by $\sim$0.18 \AA, relative to other adatoms.
Calculations for slabs with different thickness and for the 5$\times$5
reconstruction geometry show that as the slab becomes larger, the
displacements of these two adatoms decreases. We have also explored the
type of slab used in Ref.[\ref{stich}] and [\ref{brommer}] consisting
of two reconstructed surfaces. In this case only one of the adatoms in
question is $\sim$0.2 \AA\@ higher than other adatoms.  Moreover, we
find an energetically equivalent configuration where the relative
heights of adatoms on the faulted region of the unit cell are nearly
equal and the features in the EDOS near the Fermi level shift only
slightly.  This difference in the clean surfaces is however of no
significance for the hydrogenated surfaces we study, as the relaxation
occurring after chemisorption reduces the differences between adatoms
(in the :19H surface), as discussed in the text.
\bibitem{comment2} For completeness, we have also studied a slab
capped by two reconstructed surfaces, one on each side, where the 19
DBs ($\times$2) have been saturated with hydrogen. After relaxation, no
significant differences have been found between this and the model
reported in the text, consisting of six Si layers plus H-adatoms. Since
the six-layer model has two more layers than the slab with the two
reconstructed surfaces, the atoms in the former system are displaced in
general between 0.01 and 0.04 \AA\@ respect to those in the two surface
system, as one would expect for a smaller slab system.
\end{references}
 
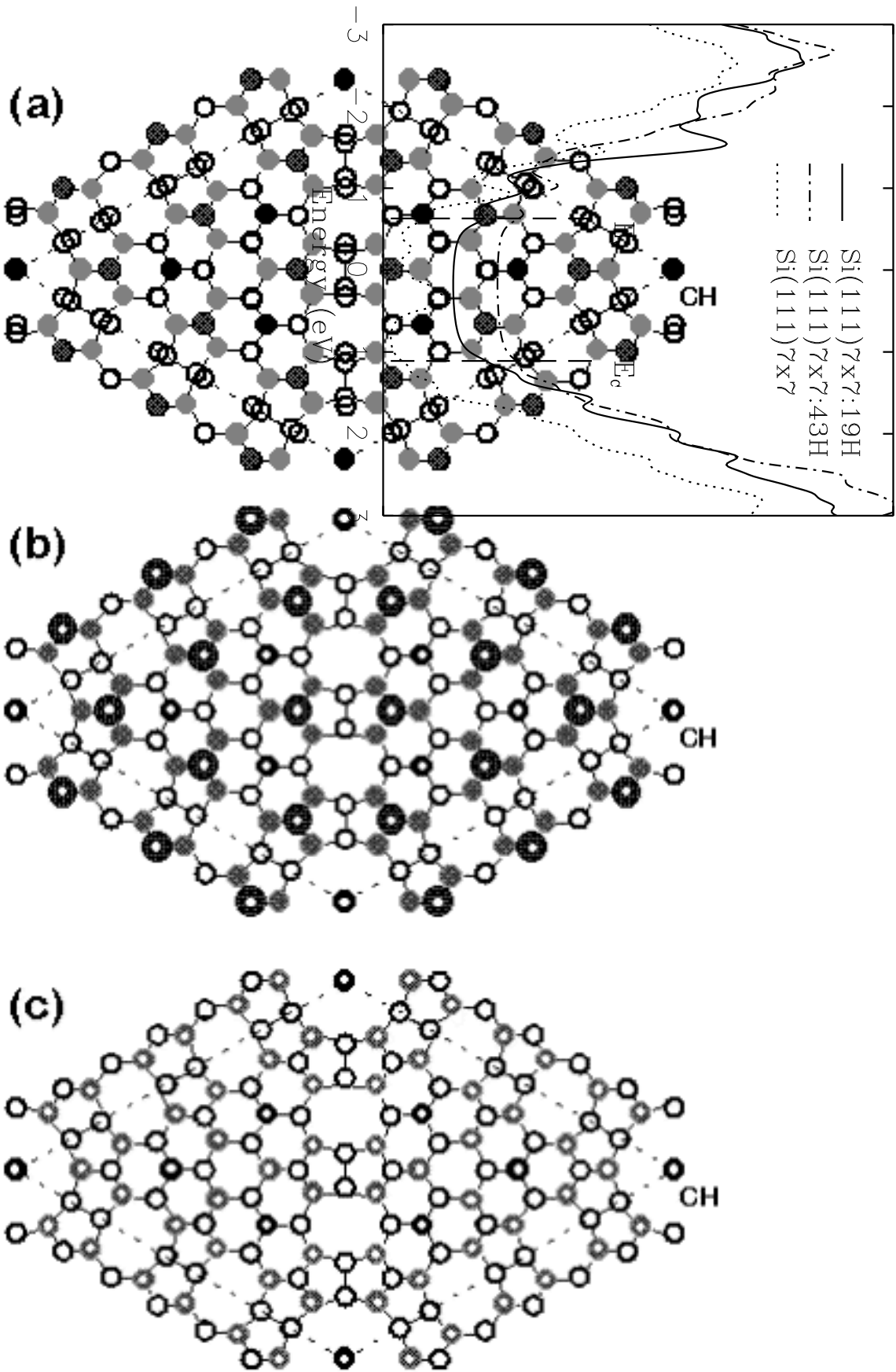
\end{document}